\documentclass[aps,prb,twocolumn,showpacs,superscriptaddress]{revtex4}

\usepackage{graphicx}% Include figure files
\usepackage{dcolumn}% Align table columns on decimal point
\usepackage{bm}% bold math
\usepackage{amssymb}

\begin{document}

\title{Electronic entanglement via quantum Hall interferometry in analogy to an optical method}

\author{Diego Frustaglia}
\email{frustaglia@us.es}
\affiliation{Departamento de F\'{\i}sica Aplicada II, Universidad de Sevilla, E-41012 Sevilla, Spain}
\author{Ad\'{a}n Cabello}
\email{adan@us.es}
\affiliation{Departamento de F\'{\i}sica Aplicada II, Universidad de Sevilla, E-41012 Sevilla, Spain}

\date{\today}

%%%%%%%%%%%%%%%%%%%%%%%%%%%%%%%%%%%%%%%%%%%%%%%%%%%%%%%%%%%%%%%%%%%

%This version: 20 October 2009

%%%%%%%%%%%%%%%%%%%%%%%%%%%%%%%%%%%%%%%%%%%%%%%%%%%%%%%%%%%%%%%%%%%

\begin{abstract}
We present an interferometric scheme producing orbital
entanglement in a quantum Hall system upon electron-hole pair
emission via tunneling. The proposed setup is an electronic version
of the optical interferometer proposed by Cabello \emph{et al.}
[Phys. Rev. Lett. {\bf 102}, 040401 (2009)], and is feasible with
present technology. It requires single-channel propagation and a
single primary source. We discuss the creation of entanglement and
its detection by the violation of a Bell inequality.
\end{abstract}

%%%%%%%%%%%%%%%%%%%%%%%%%%%%%%%%%%%%%%%%%%%%%%%%%%%%%%%%%%%%%%%%%%%

%Electronic transport in mesoscopic systems

%73.43.Fj
%Novel experimental methods; measurements (in Quantum Hall effects)
%73.43.Cd,
%Theory and modeling (in Quantum Hall effects)

\maketitle

%%%%%%%%%%%%%%%%%%%%%%%%%%%%%%%%%%%%%%%%%%%%%%%%%%%%%%%%%%%%%%%%%%%

\section{Introduction}

Experimental progress in quantum information
requires reliable sources of entanglement. In quantum optics,
spontaneous parametric down conversion is a natural source of
polarization-entangled photons \cite{KMWZSS95} and can be used to
produce energy-time entangled photons after postselection.
\cite{F-PRL89} These sources and the existence of efficient methods
for distributing photons explain the success of quantum optics for
long-distance quantum communication.

On the other hand, solid-state nanostructures offer advantages for
the local processing of quantum information. This has provoked a
major scientific effort towards the development of quantum
electronics. Specifically, there is a research program for
translating optical technologies which have already proved their
applicability for quantum information processing into the realm of
quantum electronics. That includes the development of an
electronic Mach-Zehnder interferometer, \cite{JCSHMS-Nature03}
several implementations \cite{e-HT, NOCHMU-Nature07} of
electronic Hanbury Brown-Twiss interferometers \cite{HT56} and, more
recently, the proposal \cite{GFTF-PRB06} of an electronic
Hong-Ou-Mandel interferometer. \cite{HOM87}

In this Rapid Communication we take a further step in this program 
and present a source of electronic entanglement. This is inspired by
a recent photonic interferometer, originally aimed for 
the production and detection of energy-time and time-bin 
entanglement, \cite{CRVdMM-08} after noticing that the same
scheme can be used to create orbital entanglement by
a suitable redefinition of the postselective local measurements. 
Here, we show that all topological constraints from the 
optical setup---the basis of its working principle---can be satisfied and 
the problems derived from fermionic statistics can be overcome by making use of
the last developments in quantum Hall physics. \cite{NOCHMU-Nature07}
The detection procedure is based on the measurement of zero-frequency
current-noise correlators in the tunneling regime. 
Moreover, the setup presents some distinguishing features
over previous proposals requiring either two propagating channels
\cite{BEKV-PRL03} or two sources: \cite{SSB-PRL04, SB-PRB05} 
it requires, instead, a single channel and a single tunnel barrier
as a source of correlated electron-hole pairs.

%%%%%%%%%%%%%%%%%%%%%%%%%%%%%%%%%%%%%%%%%%%%%%%%%%%%%%%%%%%%%%%%%%%

\begin{figure}[t!]
\includegraphics[width=0.90 \columnwidth,clip]{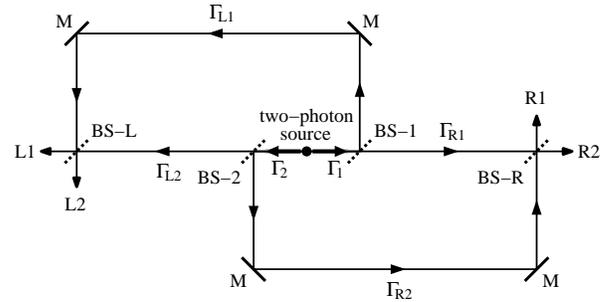}
\caption{Optical interferometer introduced in Ref.~\onlinecite{CRVdMM-08}. } 
\label{fig-1}
\end{figure}

%%%%%%%%%%%%%%%%%%%%%%%%%%%%%%%%%%%%%%%%%%%%%%%%%%%%%%%%%%%%%%%%%%%

\section{Optical interferometer}

We start by reviewing the
interferometer introduced in Ref.~\onlinecite{CRVdMM-08} (see
Fig.~\ref{fig-1}). A source simultaneously emits two photons in
opposite directions: photon 1 to the right (along path $\Gamma_1$)
and photon 2 to the left (along path $\Gamma_2$). After meeting beam
splitter BS-1 (BS-2), photon 1 (2) splits into a pair of paths
$\Gamma_{\rm R1}$ and $\Gamma_{\rm L1}$ ($\Gamma_{\rm R2}$ and
$\Gamma_{\rm L2}$). Path $\Gamma_{\rm R1}$ ($\Gamma_{\rm R2}$) takes
photon 1 (2) to the right side of the interferometer for detection,
while path $\Gamma_{\rm L1}$ ($\Gamma_{\rm L2}$) does likewise in
the left side.

The complete two-photon state emitted from BS-1 and BS-2 is a
coherent superposition of four possible paths combinations
represented by kets $| \Gamma_{\rm (L,R)1},\Gamma_{\rm (L,R)2}
\rangle$, with the first site for photon 1 and the second for photon
2. It consists of two contributions in which one photon flies off to
the right and the other one to the left ($| \Gamma_{\rm
R1},\Gamma_{\rm L2} \rangle$ and $| \Gamma_{\rm L1},\Gamma_{\rm R2}
\rangle$), and two contributions in which both photons fly off to
the same side ($| \Gamma_{\rm R1},\Gamma_{\rm R2} \rangle$ and
$|\Gamma_{\rm L1},\Gamma_{\rm L2} \rangle$). Photons 1 and 2 are not
entangled with each other. However, their state is not separable
when rewritten on a left-right bipartition basis, owning both
standard mode entanglement (orbital-mode or
path entanglement in this case) and occupation-number entanglement
(i.e., coherent superposition of terms with different local
occupation number). \cite{OccNumEnt} The orbital entanglement [i.e.,
the entanglement between left ($\Gamma_{\rm L1}$ , $\Gamma_{\rm L2}$)
and right ($\Gamma_{\rm R1}$, $\Gamma_{\rm R2}$) propagating
channels] can be postselected from the total state by coincidence
measurements at both sides of the interferometer. This keeps only
that part of the two-photon state with one photon on each side of
the interferometer: events in which two photons arrive in the same
side are simply rejected. The postselected state corresponds
to a coherent superposition of $|\Gamma_{\rm L1},\Gamma_{\rm R2}
\rangle$ and $| \Gamma_{\rm R1},\Gamma_{\rm L2} \rangle$. The
additional beam splitters BS-L and BS-R produce a local mixing
required for detecting entanglement between left and right outgoing
channels via the violation of Bell inequalities. \cite{bell-ineq}

An electronic analog of this photonic setup does not require an
explicit rejection of double-click events on each side if, instead
of measuring the times of detection---something difficult in
electronic systems---, one measures zero-frequency current-noise cross
correlations in the tunneling regime. \cite{SSB-PRL04, SSB-PRL03}

The purpose of the interferometer in Ref.~\onlinecite{CRVdMM-08} was to solve a
fundamental deficiency in the Franson's Bell experiment \cite{F-PRL89}
based on energy-time and time-bin entanglement,
identified by Aerts {\em et al.} \cite{AKLZ-PRL99-PRL01}.
Interestingly, the Franson's interferometer (including its electronic analog)
cannot be used to produce orbital entanglement, since the postselection in
Franson's scheme requires communication between the local parties
and cannot be avoided by a local redefinition of the observables.

%%%%%%%%%%%%%%%%%%%%%%%%%%%%%%%%%%%%%%%%%%%%%%%
\begin{figure}[t!]
\includegraphics[width=\columnwidth,clip]{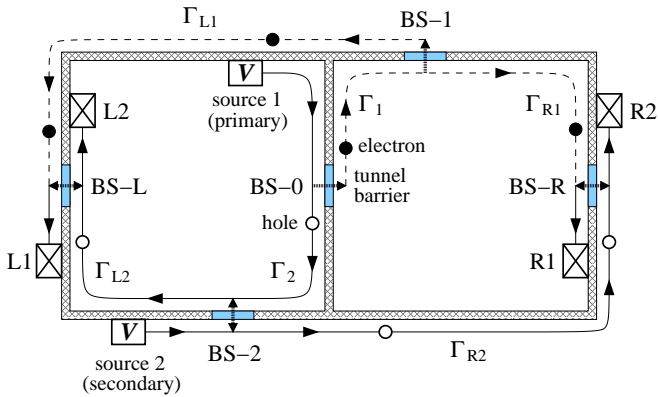}
\caption{Electronic analogue of the interferometer of
Fig.~\ref{fig-1} on a quantum Hall setup. Full (noiseless) electron
streams---redefined vacuum (see text)---are represented by solid
lines. Dashed lines correspond to empty electron channels.}
\label{fig-2}
\end{figure}
%%%%%%%%%%%%%%%%%%%%%%%%%%%%%%%%%%%%%%%%%%%%%%%

\section{Electronic interferometer}

Figure~\ref{fig-2} represents the
electronic implementation of the interferometer of Fig.~\ref{fig-1}
on a quantum Hall system. The resulting device is feasible nowadays
with modern experimental techniques. \cite{NOCHMU-Nature07} Though
geometrically similar to that of Fig.~\ref{fig-1}, the electronic
version has some singular features as a consequence of the fermionic
nature of the carriers. Electrons propagate coherently along
single-mode edge channels from sources 1 and 2 (subject to equal
voltages $V$) to drains L1, L2, R1, and R2 (connected to earth). On
their way, the electrons find a series of electrically controlled
quantum point contacts acting as beam splitters (BS-$n$, with $n =
0, 1, 2, {\rm L}, {\rm R}$). The BS-0 is set to be low transmitting
(tunnel barrier). An electron propagating from \emph{primary} source
1 can tunnel through BS-0 to the right side of the barrier, leaving
a hole in the Fermi sea on the left side. So, BS-0 behaves as an
electron-hole pair emitter (as discussed in Ref.~\cite{BEKV-PRL03},
with the difference that here we consider single---instead of
double---channel propagation \cite{note-1}). After emission, each
member of the electron-hole pair splits independently into a pair of
paths at BS-1 and BS-2, respectively, as discussed above. Path
entanglement can be observed by zero-frequency current-noise cross
correlations, which were shown \cite{SSB-PRL04, SSB-PRL03, B-EFS05}
to be equivalent to coincidence measurements in the tunneling
regime. The \emph{secondary} source 2 is not directly involved in
the production of entanglement itself: its role is to eliminate the
undesired current-noise correlations that otherwise would be generated
at BS-2,
masking the signal originated from the creation of electron-hole
pairs at BS-0. Moreover, note that the resulting entanglement is not
exactly between electrons on the one side and holes on the other
side (in contrast to previous proposals \cite{BEKV-PRL03,
SSB-PRL04}), since both right and left propagating excitations can
be either electronlike or holelike due to the combined action of BS-1
and BS-2. The BS-L and BS-R produce a controllable local mixing of
right and left propagating channels, as discussed below.

%%%%%%%%%%%%%%%%%%%%%%%%%%%%%%%%%%%%%%%%%%%%%%%%%%%%%%%%%%%%%%%%%%%

\section{Entanglement production}

We start by introducing the
uncorrelated state injected from sources $n= 1, 2$ as
\begin{equation}
|\Psi_{\rm in} \rangle=\prod_{0 < \varepsilon < eV} a_1^\dag(\varepsilon) a_2^\dag(\varepsilon) |0\rangle,
\label{Psi-in}
\end{equation}
where the operator $a_m^\dag(\varepsilon)$ excites an electron
towards BS-0 ($m=1$) and BS-2 ($m=2$) with energy $\varepsilon$ on
an energy window $eV$ above the Fermi sea $|0 \rangle$. Upon
tunneling of electrons from source 1 through BS-0, the initial state
(\ref{Psi-in}) scatters as
\begin{equation}
|\Psi' \rangle=\prod_{ \varepsilon} \left [t_0 b_1^\dag(\varepsilon) + r_0 b_2^\dag(\varepsilon) \right ] a_2^\dag
(\varepsilon) |0\rangle,
\label{Psi-inter}
\end{equation}
where $t_0$ and $r_0$ are the scattering amplitudes at BS-0
($T_0=|t_0|^2 \ll R_0=|r_0|^2$), and $b_n^\dag(\varepsilon)$ excites
a propagation mode from BS-0 toward BS-$n$ ($n=1,2$). Expanding
Eq.~(\ref{Psi-inter}) up to first order in $t_0$ and using
$b_2(\varepsilon)b_2^\dag(\varepsilon) |0\rangle = |0\rangle$ (close
to what is done in Ref.~\cite{SSB-NJP05}), we find
\begin{equation}
|\Psi' \rangle \approx \left [1- t_0 \int_0^{eV} {\rm
d}\varepsilon'~ b_2(\varepsilon') b_1^\dag(\varepsilon')\right ]
\prod_{ \varepsilon} b_2^\dag(\varepsilon) a_2^\dag(\varepsilon)
|0\rangle. \label{Psi-inter-2}
\end{equation}
The integral term in Eq.~(\ref{Psi-inter-2}) corresponds to the emission
(with probability $T_0 \ll 1$) of an electron-hole pair packet from
BS-0, where the electron ($b_1^\dag$) propagates to the right and
the hole ($b_2$) to the left (see Fig.~\ref{fig-2}). The hole
appears as an excitation out of a full stream of particles toward
BS-2 represented by $\prod_{ \varepsilon} b_2^\dag(\varepsilon)
|0\rangle$. The electrons emitted from source 2 ($a_2^\dag$) do not
play any role in the generation of the electron-hole pair. Their
relevance is proved only after scattering at BS-2, as we see next.
Upon scattering at BS-1 and BS-2 (with amplitudes $t_1,r_1$ and $t_2,r_2$, 
respectively),
the intermediate state (\ref{Psi-inter-2}) evolves into
\begin{equation}
|\Psi_{\rm out} \rangle = |\bar{0}\rangle + |\bar{\Psi}\rangle,
\end{equation}
where
\begin{eqnarray}
\label{Psi-bar} |\bar{\Psi} \rangle &=& t_0 e^{i(\phi_1-\phi_2)}
\int_0^{eV} {\rm d}\varepsilon'~ [ t_1 t_2^*C_{\rm L1}^\dag(\varepsilon') C_{\rm R2}(\varepsilon') \nonumber \\ 
&-& r_1 r_2^* C_{\rm L2}(\varepsilon') C_{\rm R1}^\dag(\varepsilon')
+ t_1 r_2^* C_{\rm L1}^\dag(\varepsilon') C_{\rm L2}(\varepsilon') \nonumber \\
&+& r_1 t_2^* C_{\rm R1}^\dag(\varepsilon') C_{\rm R2}(\varepsilon') ]
|\bar{0}\rangle
\end{eqnarray}
describes an electron-hole excitation out of a redefined vacuum
$|\bar{0}\rangle = \prod_\varepsilon^{eV} C_{\rm
L2}^\dag(\varepsilon) C_{\rm R2}^\dag(\varepsilon) |0\rangle$. Here,
$C_n^\dag$ ($C_n$) creates an electron (hole) propagating towards
terminal $n=$L1, L2, R1, or R2 (when BS-L and BS-R are
\emph{closed}) and $\phi_m$ is the phase acquired by an electron
along path $\Gamma_m$ with $m= 1, 2$. The redefined vacuum corresponds to
a \emph{noiseless} stream of electrons emitted from BS-2 toward
terminals L2 and R2. This is only possible thanks to the
introduction of the secondary source 2, which sets the net current through
BS-2 to zero when BS-0 is closed. Otherwise, electrons from
primary source 1 alone would be scattered at BS-2 as correlated
noisy currents, masking the signatures of the electron-hole emission
at BS-0.

Having a look at $|\bar{\Psi}\rangle$ in Eq.~(\ref{Psi-bar}), and leaving aside
the specific features of electrons, we notice
that the electron-hole pair emitted from BS-0 suffers from an
evolution wholly analogous to the one described in
Ref.~\cite{CRVdMM-08} for photon pairs (as discussed above).

The first two terms within brackets in Eq.~(\ref{Psi-bar}) show
a coherent
superposition of an electron and a hole traveling, alternately,
toward opposite sides of the interferometer along different paths.
This is the part of the state we are interested in, corresponding to
a pair of ``orbital'' qubits, \cite{note-qubits} which can be
entangled depending on the relative weights given by the scattering
amplitudes at BS-1 and BS-2.
The entanglement of a normalized two-qubit pure state $| \Psi
\rangle$ can be quantified by the concurrence $0 \le {\cal C}(\Psi)=
|\langle \Psi| \tilde{\Psi} \rangle| \le 1$, \cite{W-PRL98} where $|
\tilde{\Psi} \rangle = \sigma_y \otimes \sigma_y | \Psi^* \rangle$
is the time reverse of $ | \Psi \rangle$ (with $\sigma_y$ the second
Pauli matrix), ${\cal C}(\Psi)=0$ for separable states (no
entanglement), and ${\cal C}(\Psi)=1$ for Bell states (maximal
entanglement). Applying this to 
the first two terms in
Eq.~(\ref{Psi-bar}), after normalization, we obtain
\begin{equation}
{\cal C}= 2 \frac {\sqrt{T_1T_2R_1R_2}}{T_1T_2+R_1R_2},
\label{conc}
\end{equation}
where $T_1$ ($R_1$) and $T_2$ ($R_2$) are the transmission
(reflection) probabilities at BS-1 and BS-2, respectively. Maximal
entanglement is achieved whenever $T_1 T_2 = R_1 R_2$.

The last two terms within brackets in
Eq.~(\ref{Psi-bar}), instead, correspond to a
particle and a hole traveling both either to the right or to the
left side of the interferometer. This part of the state (subject to
occupation-number entanglement only) shall be filtered out during
measurement.

\section{Entanglement detection}

At this point, we can drop out of the
electron-hole picture introduced above, which was only a convenient
frame for revealing the process of entanglement production. From now
on we work within the standard electron picture, which simplifies the
description of the detection procedure. In the tunneling regime, entanglement
can be detected
via the violation of Bell-like inequalities \cite{bell-ineq} constructed
upon the measurement of zero-frequency current-noise cross correlations
defined as
\begin{equation}
S_{ij}=\lim_{\cal{T} \rightarrow \infty} \frac {h\nu} {{\cal{T}}^2} \int_0^{\cal{T}} {\rm d}t_1 {\rm d}t_2
\langle \delta I_{{\rm L}i}(t_1) \delta I_{{\rm R}j}(t_2)\rangle.
\end{equation}
This quantity correlates the time-dependent current fluctuations
$\delta I_{{\rm L}i}$ at the left terminals ($i=1, 2$) with the
fluctuations $ \delta I_{{\rm R}j}$ at the right terminals
($j=1,2$), where $\cal{T}$ is the measurement time and $\nu$ is the
density of states [a discrete spectrum is considered to ensure a
proper regularization of the current-noise correlations 
(see Ref.~\onlinecite{GFTF-PRB06})].  
The last two terms in
Eq.~(\ref{Psi-bar}) do not
contribute to $S_{ij}$, since this is a two-particle observable
demanding the presence of one particle on each side of the
interferometer for detection. Thanks to this, only 
the first two terms in
Eq.~(\ref{Psi-bar}) are postselected. At low temperatures
($kT<<eV$), the cross correlator reads \cite{B-PRL90} as
\begin{eqnarray}
S_{ij}&=&-e^3V/h |( t_{\rm L} t_{\rm R}^\dag)_{ij}|^2,
\label{s-noise}
\end{eqnarray}
where the $2 \times 2$ matrices $t_{\rm L}$ and $t_{\rm R}$ contain
the scattering amplitudes from sources 1 and 2 to terminals L1 and L2
the first one, and to R1 and R2 the second one. They satisfy $t_{\rm
L}^\dag t_{\rm L}+ t_{\rm R}^\dag t_{\rm R}=\openone$ due to
unitarity of the scattering matrix. The $S_{ij}$ turns out to be
proportional to the tunneling probability $T_0$ (i.e., $S_{ij}
\propto T_0$), meaning that any correlation signal is due to the
emission of electron-hole pairs from BS-0 alone. This is only possible
thanks to the presence of secondary source 2: otherwise,
$S_{ij}$ would be finite even for $T_0=0$, due to the undesired
correlated noise generated at BS-2. So defined, the correlator
$S_{ij}$ is proportional to the probability of joint detection of
particles in terminals $i$ and $j$ (an electron on one side and a
hole on the other). \cite{SSB-PRL04,SSB-PRL03, B-EFS05}
The presence of finite reference currents at both sides
of the interferometer (redefined vacuum $|\bar{0}\rangle$)
does not change this fact. This is because the current-noise correlators
are independent of the noiseless reference currents (an alternative
formulation based on pure tunneling currents can be used with identical
results). A Bell inequality can
be constructed upon $S_{ij}$ by defining the correlation function
\begin{eqnarray}
E= \frac {S_{11}+S_{22}-S_{12}-S_{21}}
{S_{11}+S_{22}+S_{12}+S_{21}}= \frac {{\rm tr} \left( \sigma_z
t_{\rm L} t_{\rm R}^\dag \sigma_z t_{\rm R} t_{\rm L}^\dag\right)}
{{\rm tr} \left(t_{\rm L}^\dag t_{\rm L} t_{\rm R}^\dag\ t_{\rm
R}\right)},
\end{eqnarray}
where $\sigma_z$ is the third Pauli matrix. The correlator $E$ is
explored by introducing an additional local mixing of left and right
outgoing channels. This is implemented through the beam splitters
BS-L and BS-R, as shown in Fig.~\ref{fig-2}, from which the
transmission matrices transform as $t_{\rm L} \rightarrow U_{\rm L}
t_{\rm L}$ and $t_{\rm R} \rightarrow U_{\rm R} t_{\rm R}$, where
$U_{\rm L}$ and $U_{\rm R}$ are the corresponding $2 \times 2$
unitary scattering matrices. \cite{note-phases} Hence, the
correlator $E$ transforms as
\begin{eqnarray}
E(U_{\rm L},U_{\rm R})= \frac {{\rm tr} \left(U_{\rm L}^\dag
\sigma_z U_{\rm L} t_{\rm L} t_{\rm R}^\dag U_{\rm R}^\dag \sigma_z
U_{\rm R} t_{\rm R} t_{\rm L}^\dag\right)} {{\rm tr} \left(t_{\rm
L}^\dag t_{\rm L} t_{\rm R}^\dag\ t_{\rm R}\right)},
\end{eqnarray}
from which the Bell-Clauser-Horne-Shimony-Holt (CHSH) operator is
defined as \cite{bell-ineq}
\begin{eqnarray}
{\cal E}=
E(U_{\rm L},U_{\rm R})+E(U'_{\rm L},U_{\rm R})+E(U_{\rm L},U'_{\rm R})-E(U'_{\rm L},U'_{\rm R}). \nonumber \\
\label{bell-param}
\end{eqnarray}
The studied state is entangled if the Bell-CHSH operator satisfies
$|{\cal E}| > 2$ for some configurations of matrices $\{ U_{\rm
L},U_{\rm R},U'_{\rm L},U'_{\rm R}\}$. Following
Refs.~\onlinecite{BEKV-PRL03} and \onlinecite{PR-PLA92}, we find that the
maximum possible value for the Bell-CHSH operator (\ref{bell-param})
reads as
\begin{eqnarray}
{\cal E}_{\rm max}= 2 \sqrt{1+ \frac {4 (1-\lambda_+)(1-\lambda_-)\lambda_+ \lambda_-}
{(\lambda_++\lambda_--\lambda_+^2-\lambda_-^2)^2}},
\label{E-max}
\end{eqnarray}
where
$\lambda_{+} = 1- T_0 T_1 T_2$ and $\lambda_{-} = T_0 R_1 R_2$
are the eigenvalues of the matrix product $t_{\rm R}^\dag t_{\rm R}$ up
to first order in the tunneling probability $T_0$.
We notice that Eq.~(\ref{E-max}) reduces to ${\cal E}_{\rm max} = 2 \sqrt{1 + {\cal
C}^2}$ with ${\cal C}$ the concurrence of Eq.~(\ref{conc}). This is
an expected relation for a pair of entangled qubits,
\cite{BEKV-PRL03, G-PLA91} which guarantees the accuracy of our
approach. Its meaning is straightforward in our case: whenever there
is orbital entanglement in the emitted state $|\bar{\Psi} \rangle$
of Eq.~(\ref{Psi-bar}) (${\cal C} > 0$), there is a violation of the Bell-CHSH inequality
$|{\cal E}| \le 2$.

%%%%%%%%%%%%%%%%%%%%%%%%%%%%%%%%%%%%%%%%%%%%%%%%%%%%%%%%%%%%%%%%%%%%%

\section{Conclusions} 

The production and detection of entanglement are still 
a major challenge for quantum electronics. Here we have described a  
source of orbital entanglement in an electron-hole quantum Hall systems.  
We have discussed how to use it to prepare entangled states and characterize 
entanglement and quantum nonlocality. A fundamental feature is that the 
scheme is simpler than previous proposals and seems feasible with 
present technology, so we expect that it can stimulate further experimental 
developments in electronic quantum information.

%%%%%%%%%%%%%%%%%%%%%%%%%%%%%%%%%%%%%%%%%%%%%%%%%%%%%%%%%%%%%%%%%%%%%

\acknowledgments{
We acknowledge support from the Ram\'on y Cajal program, from the
Spanish Ministry of Science and Innovation's Project No.
FIS2008-05596, and from the Junta de Andaluc\'{\i}a's Excellence
Projects No. P06-FQM-2243 and No. P07-FQM-3037.
}
%%%%%%%%%%%%%%%%%%%%%%%%%%%%%%%%%%%%%%%%%%%%%%%%%%%%%%%%%%%%%%%%%%%%%


\begin{thebibliography}{99}
\bibitem{note-1} For the sake of simplicity, we consider the spinless case only.
Spin-degenerate channels can be introduced without any significant
consequence on orbital entanglement due to separability of spin
and orbital degrees of freedom.
\bibitem{note-phases}
The $U_{\rm L}$ and $U_{\rm R}$ mixing matrices can absorb the
kinetic/magnetic phases acquired by the electrons (and holes) along
their way from BS-1 and BS-2 to drains, so we do not need to
consider them explicitly. Notice that the mixing can be performed
either by controlling the transmission amplitude of BS-L and BS-R,
by modifying the Aharonov-Bohm flux through the interferometer, or
by introducing local electric gates that modify the length of
selected paths towards the left and/or right side of the
interferometer.
\end{thebibliography}
\end{document}